\documentclass{article}

\usepackage{arxiv}
\usepackage[utf8]{inputenc}
\usepackage[T1]{fontenc}
\usepackage[numbers,sort&compress]{natbib}
\usepackage{hyperref}
\usepackage{url}
\usepackage{booktabs}
\usepackage{amsfonts}
\usepackage{amsmath,amssymb}
\usepackage{nicefrac}
\usepackage{microtype}
\usepackage{graphicx}
\usepackage{placeins}
\graphicspath{{./figures/}}

\title{Hardware--Software Co-Design for Event-Driven SNN Deployment on Low-Cost Neuromorphic FPGAs}

\author{
  Jiwoon Lee \\
  Department of Computer Engineering, \\
  Kwangwoon University \\
  Seoul, South Korea \\
  \texttt{jwlee@linux.com} \\
  \And
  Souvik Chakraborty \\
  Department of Applied Mechanics, \\
  Indian Institute of Technology Delhi, \\
  New Delhi, India \\
  Department of Nuclear, Plasma and \\
  Radiological Engineering, \\
  University of Illinois Urbana-Champaign, \\
  Urbana, IL, USA \\
  \texttt{souvik@am.iitd.ac.in} \\
  \And
  Syed Bahauddin Alam \\
  Department of Nuclear, Plasma and \\
  Radiological Engineering, \\
  University of Illinois Urbana-Champaign, \\
  Urbana, IL, USA \\
  \texttt{alams@illinois.edu} \\
  \And
  Cheolsoo Park \\
  Department of Computer Engineering, \\
  Kwangwoon University \\
  Seoul, South Korea \\
  \texttt{parkcheolsoo@kw.ac.kr}
}

\begin{document}
\maketitle

\noindent\textbf{Publication note.}
This arXiv version has been updated to incorporate the revisions made for the peer-reviewed ICONS '26 paper.
The version of record was published in the \emph{Proceedings of the International Conference on Neuromorphic Systems (ICONS '26)}, pp.~124--127.
Please refer to the published conference paper for the version of record:
\href{https://doi.org/10.1145/3822454.3822493}{doi:10.1145/3822454.3822493}.

\begin{abstract}
Low-cost FPGA boards are widely available, but deploying PyTorch-defined spiking neural networks (SNNs) on event-driven FPGA hardware remains cumbersome. We propose a deployment flow that preserves model semantics from PyTorch definition to board execution. A single exported artifact carries weights, thresholds, connectivity descriptors, and grouped time-to-first-spike (TTFS) decoding metadata, and is reused unchanged by both the software reference runner and the board runtime. A 10-class MNIST TTFS classifier on the routed 80\,MHz design achieves 87.40\% accuracy and matches the software reference on all 10{,}000 test images. The PL-only path achieves a steady-state service latency of 0.1375\,$\mu$s/image and an estimated PL-dynamic energy of 31.6\,nJ/image. We separately report host-inclusive runtime and board-level energy to avoid mixing accelerator-only and system-level scopes. Together, these results demonstrate a reproducible software-to-board path for SNN deployment on low-cost FPGA hardware.
\end{abstract}


\section{Introduction}
Digital neuromorphic processors and spike-based systems are actively studied as low-power computing platforms~\cite{painkras2013spinnaker,park2022high,park2024high}. FPGA accelerators remain attractive for research because they are inexpensive, reconfigurable, and easy to instrument, and prior work has demonstrated the value of event-driven FPGA SNN execution~\cite{neil2014minitaur}. Yet practical SNN workflows remain split between hardware-first and software-first paths. Hardware-first approaches often expose HDL-centered, clock-driven implementations that are efficient but awkward to integrate into software-defined SNN workflows. Research platforms such as Intel's Loihi and the associated Lava stack narrow this gap, but access to such systems remains more limited than access to low-cost commodity FPGA boards~\cite{davies2018loihi,intel2021lava}. Software-first approaches make model construction and evaluation easy in PyTorch-style environments, but often stop at simulation or GPU execution~\cite{paszke2019pytorch,fang2023spikingjelly,hazan2018bindsnet,mozafari2019spyketorch}. A direct deployment path that preserves model semantics from software definition to event-driven hardware remains limited, especially on low-cost FPGA platforms.

We address this gap with a single exported artifact that is reused unchanged by both the Python reference path and the FPGA runtime. We then verify prediction-level agreement over the full MNIST test set rather than relying only on aggregate accuracy.

Compared with prior digital neuromorphic processors and event-driven FPGA accelerators~\cite{park2024high,neil2014minitaur}, the emphasis here is a usable deployment framework that carries software-defined SNN models to event-driven hardware while keeping hardware/software agreement and measurement scopes explicit.

The contributions are threefold. First, the framework provides a PyTorch-aligned deployment flow in which a single artifact preserves deployed model semantics across software reference execution and board runtime without a separate board-specific conversion stage. Second, it demonstrates deterministic TTFS~\cite{park2020t2fsnn} inference on PYNQ-Z2 by matching the software reference on every image in the full 10{,}000-image MNIST~\cite{lecun1998mnist} test set, a stricter test than aggregate accuracy alone for first-spike classifiers~\cite{goltz2021fast,park2020t2fsnn}. Third, it combines software-facing usability with event-driven neuromorphic execution while separating accelerator-only from host-inclusive measurements.

\begin{figure}[t]
  \centering
  \includegraphics[width=0.98\textwidth]{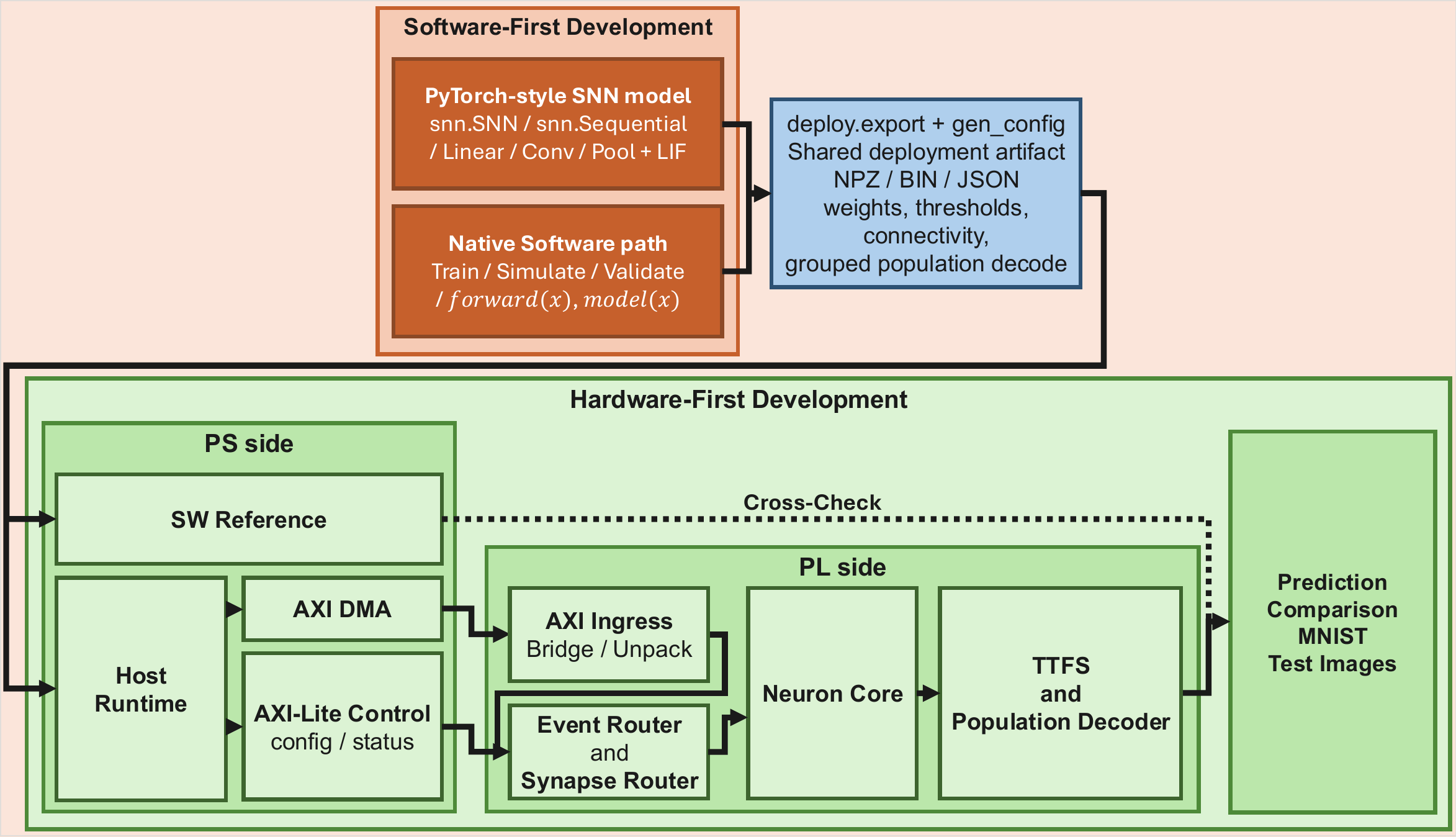}
  \caption{Overview of the proposed hardware-software co-design path. The same deployment artifact is consumed by both the software reference runner and the event-driven board runtime on PYNQ-Z2. The programmable logic implements the event router, core groups, connectivity table, grouped TTFS decoder, and timing counters.}

  \label{fig:architecture}
\end{figure}

\section{Architecture and Methods}
\subsection{Architecture}
The target platform is the PYNQ-Z2, built around a Xilinx Zynq-7020 SoC. The deployed inference design comprises an AXI-fed event router, a grouped spike-processing fabric, a connectivity table, a grouped TTFS decoder, and on-chip PL counters for latency measurement. The current RTL event-processing path directly addresses up to 2{,}048 neurons, organized as 16 groups of 128. The exported configuration format and software-side deployment flow can represent larger model descriptions, including the 4{,}890-neuron and 843{,}776-packed-synapse configuration, but the present routed XC7Z020 bitstream is limited by on-chip BRAM capacity and directly validates the 2{,}048-neuron event-processing fabric reported here. The routed implementation meets timing at 80\,MHz and is constrained primarily by on-chip memory. Table~\ref{tab:impl} and Figure~\ref{fig:architecture} summarize the design.

\subsection{Software interface and deployment}
Model-side modules follow \texttt{nn.Module}-style construction, and the runtime exposes a PyTorch-compatible module invocation interface~\cite{paszke2019pytorch}. Export generates a single deployment artifact containing weights, thresholds, connectivity descriptors, and grouped decoding metadata. The same artifact is consumed unchanged by the software reference runner and the board runtime. Table~\ref{tab:stack} summarizes the deployed subset and the corresponding PyTorch-style abstractions.

\subsection{Board validation and measurement protocol}
Board-side evaluation uses a 10-class MNIST TTFS classifier whose deployment artifact preserves weights, thresholds, and grouped decoding rules across software reference execution and board inference. The deployed classifier uses 150 output neurons arranged as 10 class groups of 15 neurons each~\cite{goupy2024neuronal,diehl2015unsupervised}, and the decoded label is obtained from a grouped TTFS readout over these class-specific neuron populations~\cite{park2020t2fsnn}. For each test example, the class decoded from the PL output stream is compared with the software TTFS reference generated from the same artifact. In the main reported run, all 10{,}000 board predictions match the software reference, confirming end-to-end preservation of the deployed decision rule.

\begin{table}[t]
  \caption{Post-route resource utilization and timing summary of the deployed 80\,MHz design on PYNQ-Z2. The design is BRAM-limited, while logic and DSP resources retain substantial headroom.}
  \label{tab:impl}
  \centering
  \scriptsize
  \begin{tabular}{lcc}
    \toprule
    Resource & Used / available & Util. (\%) \\
    \midrule
    Slice LUTs & 14{,}552 / 53{,}200 & 27.35 \\
    Slice registers & 9{,}955 / 106{,}400 & 9.36 \\
    DSP48E1 & 17 / 220 & 7.73 \\
    Block RAM tiles & 140 / 140 & 100.00 \\
    \midrule
    Limiting resource & \multicolumn{2}{c}{On-chip BRAM capacity} \\
    Clock frequency & \multicolumn{2}{c}{80\,MHz} \\
    Worst setup slack & \multicolumn{2}{c}{+0.147\,ns} \\
    Worst hold slack & \multicolumn{2}{c}{+0.032\,ns} \\
    \bottomrule
  \end{tabular}
\end{table}

\begin{table}[t]
  \caption{Software-to-board deployment workflow and corresponding PyTorch-style abstractions used in this work. The same model definition, export artifact, and invocation interface are used for both software reference execution and board execution.}
  \label{tab:stack}
  \centering
  \scriptsize
  \begin{tabular}{p{0.16\linewidth}p{0.24\linewidth}p{0.22\linewidth}p{0.23\linewidth}}
    \toprule
    Stage & Framework interface & PyTorch-style counterpart & Role in this work \\
    \midrule
    Model definition & \texttt{snn.SNN}, \texttt{snn.Sequential}, \texttt{snn.Linear}, \texttt{snn.LIF} & \texttt{nn.Module}, \texttt{nn.Sequential}, \texttt{nn.Linear} + activation stage & Defines the reported 784-to-150 TTFS classifier. \\
    Artifact export & \texttt{deploy.export}, \texttt{deploy.gen\_config} & Export companion to the module graph & Produces one shared deployment artifact. \\
    Runtime inference & \texttt{SNNAccelerator} runtime & PyTorch-compatible module invocation & Executes the shared artifact through the reference runner and board driver. \\
    \bottomrule
  \end{tabular}
\end{table}

Each reported result uses the full 10{,}000-image MNIST test set at the implemented 80\,MHz PL clock. We distinguish two latency scopes. \emph{Accelerator latency} is measured from on-chip cycle counters and captures only PL execution, including first-output-spike latency and steady-state service latency after pipeline fill. \emph{Host-inclusive validation latency} is the wall-clock time per image in the validation profile and includes software reference evaluation, spike packing, DMA orchestration, polling, and readback. Figure~\ref{fig:scope} visualizes the dominant host-inclusive validation-path latency terms together with the PL-only timing summary.

For software baselines, we evaluate dense grouped-neuron executions of the same exported 784-to-150 classifier on an RTX~3080 GPU and on a CPU using a NumPy implementation in FP32 and INT8 modes. These baselines reuse the same learned parameters and grouped readout but execute them as dense software kernels rather than event-level TTFS runtimes. PL-only FPGA latency is therefore compared against GPU kernel-only and CPU compute-only paths.

Energy is handled similarly. GPU kernel energy is estimated from the device's active-minus-idle power over the benchmark window, whereas CPU energy is omitted because no comparable runtime power trace is available. For the FPGA, the cross-platform comparison uses a PL-dynamic estimate derived from the Xilinx power-analysis flow, while XADC-derived board activity is reported separately as a whole-system view~\cite{amd2025ug907,amd2022ug480}.

\section{Measurement Results}
\subsection{Cross-Platform Comparison}
Table~\ref{tab:main} summarizes the main cross-platform result. The FPGA achieves the same 87.40\% MNIST accuracy as the software TTFS reference and matches the software reference on all 10{,}000 test images. The 12-cycle first-spike latency captures pipeline fill, whereas the 11-cycle steady-state service latency characterizes the response after fill. At 80\,MHz, the corresponding PL-only service latency is 0.1375\,$\mu$s/image, or 7.27\,Mimages/s. Relative to topology-matched software baselines, the GPU INT8 kernel is 1.79$\times$ slower and about 933$\times$ higher in dynamic energy per image, while the CPU INT8 path is 488$\times$ slower.

The full-test-set prediction match confirms that the shared deployment artifact preserves the deployed decision rule across the Python TTFS reference runner and the FPGA PL output stream. The GPU and CPU rows are parameter- and readout-matched dense software baselines: they reuse the exported weights and grouped readout but do not implement the same event-level TTFS runtime. Their small accuracy differences should therefore be interpreted as baseline-execution differences, not as mismatches in the FPGA deployment path.

\begin{table}[t]
  \caption{Scope-separated performance of the deployed 10-class MNIST classifier. FPGA latency is PL-only and FPGA energy is a PL-dynamic estimate; GPU latency and energy are kernel-only estimates, and CPU latency is compute-only with energy not measured. GPU and CPU rows are dense software baselines rather than event-level TTFS hardware replicas.}
  \label{tab:main}
  \centering
  \scriptsize
  \begin{tabular}{lrrrr}
    \toprule
    Platform & Acc. (\%) & Latency ($\mu$s/img) & Throughput (img/s) & Energy (nJ/img) \\
    \midrule
    FPGA (Ours, PL-only) & 87.40 & 0.1375 & $7.27\times10^6$ & 31.6 \\
    GPU FP32 & 87.69 & 0.2874 & $3.48\times10^6$ & $5.94\times10^4$ \\
    GPU INT8 & 87.70 & 0.2461 & $4.06\times10^6$ & $2.95\times10^4$ \\
    CPU FP32 & 87.69 & 88.99 & $1.12\times10^4$ & N/A \\
    CPU INT8 & 87.70 & 67.03 & $1.49\times10^4$ & N/A \\
    \bottomrule
  \end{tabular}
\end{table}

\subsection{Host-Inclusive Runtime and Energy}
\begin{figure}[t]
  \centering
  \includegraphics[width=0.9\linewidth]{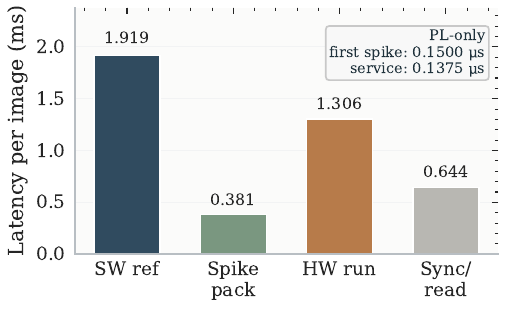}
  \caption{Host-inclusive validation-path latency breakdown. The bars separate software reference evaluation, spike packing, hardware execution/orchestration, and synchronization/readback, while the callout reports PL-only first-spike and steady-state service latencies.}

  \label{fig:scope}
\end{figure}

The host-inclusive validation profile yields a different latency breakdown. In this profile, the measured wall-clock time includes software reference evaluation, spike packing, hardware execution and orchestration, polling, and readback, and should therefore be interpreted as validation-loop latency rather than accelerator service latency. Under this profile, the wall-clock time is 4.25\,ms/image, whereas the PL-only steady-state service latency is 0.1375\,$\mu$s/image. Software reference evaluation accounts for 1.919\,ms/image, spike packing for 0.381\,ms/image, hardware execution plus orchestration for 1.306\,ms/image, and synchronization/readback for the remaining 0.644\,ms/image. Figure~\ref{fig:scope} breaks down these host-inclusive terms while retaining the PL-only first-spike and service latencies in the callout.

The same distinction appears in the energy results. The PL-dynamic estimate is 31.6\,nJ/image, whereas an XADC-based board-level energy estimate is 47.78\,mJ/image because it includes processor, memory, and software activity across the full system window. For accelerator comparison, the former is the more appropriate quantity; the latter remains relevant for deployment-level budgeting.

\subsection{Input Spike-Drop Stress and Repeatability}
Controlled input-spike dropping causes gradual degradation rather than an abrupt failure mode. Hardware TTFS accuracy decreases from 87.40\% with no spike drop to 86.31\% at 25\% spike drop and 82.38\% at 50\% spike drop, then declines to 69.74\% at 75\% spike drop.

Across five repeated runs, no prediction mismatches were observed across 50{,}000 image-run pairs, and hardware TTFS accuracy remained 87.40\% in every run. The 4.25\,ms/image result above corresponds to the host-inclusive validation profile used for the Figure~\ref{fig:scope} breakdown. In a separate embedded-host runtime profile used only to assess run-to-run stability, the wall-clock latency averages 56.77\,ms/image with a standard deviation of 0.20\,ms/image; this profile includes additional host-side overheads and is not used as the accelerator-service comparison.

\begin{figure}[t]
  \centering
  \includegraphics[width=1\linewidth]{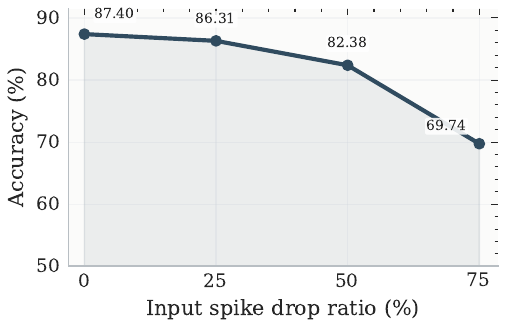}
  \caption{Input-spike drop tolerance of the deployed classifier. The curve shows gradual degradation under controlled spike-drop stress up to a 75\% drop ratio.}

  \label{fig:robustness}
\end{figure}

\FloatBarrier

\section{Discussion and Limitations}
The evaluated system provides low PL-only service latency and prediction-level traceability between the software reference and the board. This traceability is supported by the shared deployment artifact and full-test-set prediction matching, while the scope-separated timing and energy measurements facilitate interpretation and comparison. The evaluation also highlights the main scaling constraints. The reported PL-dynamic energy numbers are tool-based estimates rather than external rail measurements~\cite{amd2025ug907}. The current bitstream saturates BRAM at 140/140 tiles, so larger dense networks remain limited primarily by on-chip memory and by the effective fan-out of the current synapse-routing structure on the XC7Z020. The board-validated workload is a linear TTFS classifier. Future work will target deeper and convolutional models, with particular focus on synapse-routing schemes that increase effective fan-out under the BRAM limits of the XC7Z020.

\section{Conclusion}
We presented a hardware-software co-design flow for event-driven SNN deployment on low-cost FPGA hardware. On PYNQ-Z2, the deployed 80\,MHz implementation matches the software decision rule across the full 10{,}000-image MNIST test set while delivering 0.1375\,$\mu$s/image PL-only service latency and 31.6\,nJ/image estimated PL-dynamic energy. Overall, the single-artifact path improves the reproducibility of software-to-board SNN evaluation on low-cost neuromorphic FPGA hardware.

\section*{Code Availability}
The source code for the proposed framework and FPGA accelerator is publicly available at \url{https://github.com/ljiwoon/Event-Driven-Spiking-Neural-Network-Accelerator-for-FPGA}.

\section*{Acknowledgments}
This work was supported by the National Research Foundation of Korea (NRF) grant funded by the Ministry of Education (RS-2025-25403835) and the Institute of Information \& Communications Technology Planning \& Evaluation (IITP) grant funded by the Korea government (MSIT) (IITP-2025-RS-2022-00156225).

\bibliographystyle{unsrtnat}
\bibliography{references}

@article{painkras2013spinnaker,
  author = {Painkras, Eustace and Plana, Luis and Garside, Jim and Temple, Steve and Galluppi, Francesco and Patterson, Cameron and Lester, David and Brown, Andrew D. and Furber, Steve B.},
  title = {SpiNNaker: A 1-{W} 18-Core System-on-Chip for Massively Parallel Neural Network Simulation},
  journal = {IEEE Journal of Solid-State Circuits},
  volume = {48},
  number = {8},
  pages = {1943--1953},
  year = {2013},
  doi = {10.1109/JSSC.2013.2259038}
}

@article{neil2014minitaur,
  author = {Neil, Daniel and Liu, Shih-Chii},
  title = {Minitaur, an Event-Driven FPGA-Based Spiking Network Accelerator},
  journal = {IEEE Transactions on Very Large Scale Integration (VLSI) Systems},
  volume = {22},
  number = {12},
  pages = {2621--2628},
  year = {2014},
  doi = {10.1109/TVLSI.2013.2294916}
}

@article{davies2018loihi,
  author = {Davies, Mike and Srinivasa, Narayan and Lin, Tsung-Han and Chinya, Gautham and Cao, Yongqiang and Choday, Sriharsha and Dimou, George and Joshi, Prasad and Imam, Nabil and Jain, Shweta and Liao, Yuchen and Lin, Chung-Kuan and Lines, Andreas and Liu, Ruokun and Mathaikutty, Deepak and McCoy, Steve and Paul, Arnab and Tse, Jonathan and Venkataramanan, Gururaj and Weng, Yat-Hang and Wild, Andreas and Yang, Yoon and Wang, Hong},
  title = {Loihi: A Neuromorphic Manycore Processor with On-Chip Learning},
  journal = {IEEE Micro},
  volume = {38},
  number = {1},
  pages = {82--99},
  year = {2018},
  doi = {10.1109/MM.2018.112130359}
}

@misc{intel2021lava,
  author = {{Intel}},
  title = {Lava Software Framework},
  year = {2021},
  howpublished = {\url{https://lava-nc.org/}},
  note = {Accessed: 2026-04-07}
}

@inproceedings{park2024high,
  title={High-Density Digital Neuromorphic Processor with High-Precision Neural and Synaptic Dynamics and Temporal Acceleration},
  author={Park, Jongkil and Jeong, YeonJoo and Kim, Jaewook and Lee, Suyoun and Kwak, Joon Young and Park, Jong-Keuk and Kim, Inho},
  booktitle={2024 IEEE 6th International Conference on AI Circuits and Systems (AICAS)},
  pages={322--326},
  year={2024},
  organization={IEEE}
}

@article{park2022high,
  title={High dynamic range digital neuron core with time-embedded floating-point arithmetic},
  author={Park, Jongkil and Jeong, YeonJoo and Kim, Jaewook and Lee, Suyoun and Kwak, Joon Young and Park, Jong-Keuk and Kim, Inho},
  journal={IEEE Transactions on Circuits and Systems I: Regular Papers},
  volume={70},
  number={1},
  pages={290--301},
  year={2022},
  publisher={IEEE}
}

@article{fang2023spikingjelly,
  title={Spikingjelly: An open-source machine learning infrastructure platform for spike-based intelligence},
  author={Fang, Wei and Chen, Yanqi and Ding, Jianhao and Yu, Zhaofei and Masquelier, Timoth{\'e}e and Chen, Ding and Huang, Liwei and Zhou, Huihui and Li, Guoqi and Tian, Yonghong},
  journal={Science Advances},
  volume={9},
  number={40},
  pages={eadi1480},
  year={2023},
  publisher={American Association for the Advancement of Science}
}

@article{hazan2018bindsnet,
  title={Bindsnet: A machine learning-oriented spiking neural networks library in python},
  author={Hazan, Hananel and Saunders, Daniel J and Khan, Hassaan and Patel, Devdhar and Sanghavi, Darpan T and Siegelmann, Hava T and Kozma, Robert},
  journal={Frontiers in Neuroinformatics},
  volume={12},
  pages={89},
  year={2018},
  publisher={Frontiers Media SA}
}

@article{mozafari2019spyketorch,
  title={Spyketorch: Efficient simulation of convolutional spiking neural networks with at most one spike per neuron},
  author={Mozafari, Milad and Ganjtabesh, Mohammad and Nowzari-Dalini, Abbas and Masquelier, Timoth{\'e}e},
  journal={Frontiers in Neuroscience},
  volume={13},
  pages={625},
  year={2019},
  publisher={Frontiers Media SA}
}

@article{paszke2019pytorch,
  title={Pytorch: An imperative style, high-performance deep learning library},
  author={Paszke, Adam and Gross, Sam and Massa, Francisco and Lerer, Adam and Bradbury, James and Chanan, Gregory and Killeen, Trevor and Lin, Zeming and Gimelshein, Natalia and Antiga, Luca and others},
  journal={Advances in Neural Information Processing Systems},
  volume={32},
  year={2019}
}

@article{goltz2021fast,
  title={Fast and energy-efficient neuromorphic deep learning with first-spike times},
  author={G{\"o}ltz, Julian and Kriener, Laura and Baumbach, Andreas and Billaudelle, Sebastian and Breitwieser, Oliver and Cramer, Benjamin and Dold, Dominik and Kungl, Akos Ferenc and Senn, Walter and Schemmel, Johannes and others},
  journal={Nature Machine Intelligence},
  volume={3},
  number={9},
  pages={823--835},
  year={2021},
  publisher={Nature Publishing Group UK London}
}

@article{lecun1998mnist,
  author = {{Le Cun}, Yann and Bottou, L\'{e}on and Bengio, Yoshua and Haffner, Patrick},
  title = {Gradient Based Learning Applied to Document Recognition},
  journal = {Proceedings of IEEE},
  volume = {86},
  number = {11},
  pages = {2278-2324},
  year = {1998},
  url = {http://leon.bottou.org/papers/lecun-98h},
}

@manual{amd2025ug907,
  title = {Vivado Design Suite User Guide: Power Analysis and Optimization (UG907)},
  organization = {Advanced Micro Devices, Inc.},
  year = {2025},
  note = {Version 2025.2},
  url = {https://docs.amd.com/r/en-US/ug907-vivado-power-analysis-optimization/Output-Tab}
}

@manual{amd2022ug480,
  title = {7 Series FPGAs and Zynq-7000 SoC XADC Dual 12-Bit 1 MSPS Analog-to-Digital Converter User Guide (UG480)},
  organization = {Advanced Micro Devices, Inc.},
  year = {2022},
  note = {Revision 1.11},
  url = {https://docs.amd.com/r/en-US/ug480_7Series_XADC/Reference-Inputs-VREFP-and-VREFN}
}

@article{goupy2024neuronal,
  title={Neuronal competition groups with supervised STDP for spike-based classification},
  author={Goupy, Gaspard and Tirilly, Pierre and Bilasco, Ioan Marius},
  journal={Advances in Neural Information Processing Systems},
  volume={37},
  pages={106295--106314},
  year={2024}
}

@article{diehl2015unsupervised,
  title={Unsupervised learning of digit recognition using spike-timing-dependent plasticity},
  author={Diehl, Peter U and Cook, Matthew},
  journal={Frontiers in Computational Neuroscience},
  volume={9},
  pages={149773},
  year={2015},
  publisher={Frontiers}
}

@inproceedings{park2020t2fsnn,
  title={T2FSNN: Deep spiking neural networks with time-to-first-spike coding},
  author={Park, Seongsik and Kim, Seijoon and Na, Byunggook and Yoon, Sungroh},
  booktitle={2020 57th ACM/IEEE design automation conference (DAC)},
  pages={1--6},
  year={2020},
  organization={IEEE}
}

\end{document}